\documentclass[12pt]{article}
\usepackage{psfig}

\usepackage[dvips]{color}

\newcommand{\Black}{\color [rgb]{0,0,0}}

\newcommand{\Brown}{\color [rgb]{0.4,0.1,0.1}}

\def\TL{\hfil$\displaystyle{##}$}
\def\TR{$\displaystyle{{}##}$\hfil}

\def\TT{\hbox{##}}
\def\seqalign#1#2{\vcenter{\openup1\jot
  \halign{\strut #1\cr #2 \cr}}}

\def\fixit#1{}

\def\mop#1{\mathop{\rm #1}\nolimits}

\def\overleftrightarrow#1{\vbox{\ialign{##\crcr
     $\leftrightarrow$\crcr\noalign{\kern-0pt\nointerlineskip}
     $\hfil\displaystyle{#1}\hfil$\crcr}}}

\def\lsim{\mathrel{\mathstrut\smash{\ooalign{\raise2.5pt\hbox{$<$}\cr\lower2.5pt\hbox{$\sim$}}}}}
\def\gsim{\mathrel{\mathstrut\smash{\ooalign{\raise2.5pt\hbox{$>$}\cr\lower2.5pt\hbox{$\sim$}}}}}
\def\sqr#1#2{{\vcenter{\vbox{\hrule height.#2pt
         \hbox{\vrule width.#2pt height#1pt \kern#1pt
            \vrule width.#2pt}
         \hrule height.#2pt}}}}

\def\href#1#2{#2}

\def\lbldef#1#2{\expandafter\gdef\csname #1\endcsname {#2}}
\def\eqn#1#2{\lbldef{#1}{(\ref{#1})}
\begin{equation} #2 \label{#1} \end{equation}}
\def\eqalign#1{\vcenter{\openup1\jot
    \halign{\strut\span\TL & \span\TR\cr #1 \cr
   }}}
\def\eno#1{(\ref{#1})}
\def\Re{\mop{Re}}
\def\Im{\mop{Im}}
\def\TL{\hfil$\displaystyle{##}$}
\def\TR{$\displaystyle{{}##}$\hfil}

\def\TT{\hbox{##}}
\def\seqalign#1#2{\vcenter{\openup1\jot
  \halign{\strut #1\cr #2 \cr}}}

\def\fixit#1{}

\def\mop#1{\mathop{\rm #1}\nolimits}

\def\overleftrightarrow#1{\vbox{\ialign{##\crcr
     $\leftrightarrow$\crcr\noalign{\kern-0pt\nointerlineskip}
     $\hfil\displaystyle{#1}\hfil$\crcr}}}

\def\lsim{\mathrel{\mathstrut\smash{\ooalign{\raise2.5pt\hbox{$<$}\cr\lower2.5pt\hbox{$\sim$}}}}}
\def\gsim{\mathrel{\mathstrut\smash{\ooalign{\raise2.5pt\hbox{$>$}\cr\lower2.5pt\hbox{$\sim$}}}}}
\def\sqr#1#2{{\vcenter{\vbox{\hrule height.#2pt
         \hbox{\vrule width.#2pt height#1pt \kern#1pt
            \vrule width.#2pt}
         \hrule height.#2pt}}}}

\def\href#1#2{#2}  

\def\lbldef#1#2{\expandafter\gdef\csname #1\endcsname {#2}}
\def\eqn#1#2{\lbldef{#1}{(\ref{#1})}%
\begin{equation} #2 \label{#1} \end{equation}}
\def\eqalign#1{\vcenter{\openup1\jot
    \halign{\strut\span\TL & \span\TR\cr #1 \cr
   }}}
\def\eno#1{(\ref{#1})}
\def\Re{\mop{Re}}
\def\Im{\mop{Im}}

\begin{document}
\baselineskip=15.5pt
\pagestyle{plain}
\setcounter{page}{1}
%\renewcommand{\thefootnote}{\fnsymbol{footnote}}
%--------+---------+---------+---------+---------+---------+---------+
%Title page

\begin{titlepage}

\begin{flushright}
UCB-PTH-04/21\\
LBNL-56050\\
SLAC-PUB-10580 \\
SU-ITP-04/31\\
hep-th/0408085
\end{flushright}
\vfil

\begin{center}
{\huge Relativistic D-brane Scattering } \vskip0.5cm {\huge is Extremely Inelastic}
\end{center}

%\begin{center}
%{\huge Relativistic Brane Scattering } \vskip0.5cm {\huge is Extremely Inelastic:}
% \vskip0.5cm {\huge Open String Corrections to D-brane Dynamics}
%\end{center}

\vfil
\begin{center}
{\large Liam McAllister$^{1}$ and Indrajit Mitra$^{2,3}$}
\end{center}

$$\seqalign{\span\TL & \span\TT}{
{}^1 & SLAC and Department of Physics, Stanford University \cr\noalign{\vskip-1.5\jot}
& Stanford, CA 94309/94305 \cr
{}^2 & Berkeley Center for Theoretical Physics and Department of
Physics \cr\noalign{\vskip-1.5\jot}
& University of California, Berkeley, CA 94720-7300 \cr
{}^3 & Theoretical Physics Group, Lawrence Berkeley National Laboratory \cr\noalign{\vskip-1.5\jot}
& Berkeley, CA 94720-8162
}$$
\vfil

\begin{center}
{\large Abstract}
\end{center}

\noindent

We study the effects of quantum production of open strings on the
relativistic scattering of D-branes.  We find strong corrections
to the brane trajectory from copious production of highly-excited
open strings, whose typical oscillator level is proportional to the square of the rapidity.
In the corrected trajectory, the branes rapidly coincide and remain
trapped in a configuration with enhanced
symmetry.  This is a purely stringy effect which makes
relativistic brane collisions exceptionally inelastic. We trace this
effect to velocity-dependent corrections to the open string mass,
which render open strings between relativistic D-branes surprisingly
light. We observe
that pair-creation of open strings could play an important role in
cosmological scenarios in which branes approach each other at very
high speeds.

\vfil
\begin{flushleft}
August 2004
\end{flushleft}
\end{titlepage}
\newpage
%--------+---------+---------+---------+---------+---------+---------+
%Body
\section{Introduction}
\label{Introduction}

Thought experiments involving the scattering of strings or of
D-branes provide the key to understanding certain essential
phenomena in string theory.  The discovery of strings in the
theory is perhaps the most striking case, but other examples
include the elucidation of the sizes of strings under various
conditions and the appreciation of another length-scale in the
dynamics of slow-moving D-branes.

Despite much early interest in the scattering of D-branes, certain
important aspects of the dynamics have remained unexplored.  In
particular, the simplest treatments involve parameter regimes
governed either by supergravity or by the effective worldvolume
field theory of massless open strings.  In the latter case, there
can be significant quantum corrections arising from loops of
light open strings or from pair-production of on-shell open
strings.

A key consequence of the pair-production of open strings is the
{\it{trapping}} of D-branes \cite{KLLMMS}, which we now briefly review.
Consider two Dp-branes, $p>0$, moving with a small relative velocity.  As the
branes pass each other, the masses of stretched open strings vary with time.
This leads to pair production, in a direct analogue of the
Schwinger pair-creation process for charged particles \cite{Schwinger} or strings \cite{Porrati}
in an electric field.  Because the velocities are low, the
production of stretched strings with oscillator excitations is
highly suppressed.  The resulting unexcited stretched strings
introduce an energy cost for the branes to separate; unless these
strings can rapidly annihilate, the branes will be drawn
close together.  In collisions with a nonzero impact parameter,
the brane pair carries angular momentum; in this case the branes
spiral around their center of mass, radiating closed strings, until
eventually they fall on top of each other.  The final outcome is that the open strings
trap the branes in a configuration with enhanced symmetry.
Because this process involves the production of only unexcited open strings,
it falls within the purview of effective field theory.

Our goal is to explore related processes which are not describable in
the low-energy effective field theory but which instead involve intrinsically stringy
physics.  We will show that the ultrarelativistic scattering of
D-branes is a suitable laboratory for such an investigation, as
corrections from the massive string states turn out to be
essential.  In particular, we will demonstrate that production of
highly-excited open strings generates crucial corrections to the
brane dynamics and leads to spectacular trapping of the branes
over distances which can be of order the string
length.  As we will show, these corrections are much stronger than a naive
application of effective field theory would predict; hence this is a
setting where the importance of purely stringy effects is a
surprise. The explanation of such a huge production of highly excited
strings is that these states effectively become quite light -- the mass
receives velocity dependent corrections. The fact that open string
masses are in principle velocity-dependent is well-known, but we have
not found any explicit computations of these masses in the
literature. Our result leads to a formula for the masses of open
strings between moving D-branes.

The intuition underlying this result is that in relativistic D-brane
scattering, it should be possible to pair-produce highly-excited open strings.
The density of string states at high excitation levels grows exponentially with
energy; this is the well-known Hagedorn density of states.  For this reason, even if the production of
a given excited string state is exponentially suppressed compared to production of a massless string state,
the competition of the growing and decaying exponentials will typically cause highly-excited strings
to dominate the process, in terms of both their number and their share of the total energy.
Thus, one expects pair production of a huge number of highly excited strings.  This is
indeed the case, as was first explained by Bachas in the important work
\cite{Bachas}.  Our further observation is that because the energy transferred into these open strings
can easily be comparable to the initial kinetic energy of the D-brane pair, the massive open strings
are absolutely central to the dynamics.  This means that the backreaction arising
from purely stringy effects is crucial.

We will study the effect on the dynamics of this explosive
pair-production of massive modes.  Our conclusion is that for a large range of velocities and
impact parameters, almost all the initial kinetic energy of the branes is
transferred to open strings and to closed string radiation.  After the collision
the branes are drawn together and come to rest.  In near-miss scattering events with an
impact parameter $b$, the branes revolve around their center of
mass in a roughly circular orbit whose initial radius is of order
$b$; this orbit swiftly decays via radiation of closed strings.
This is to be contrasted to the much weaker trapping of nonrelativistic
branes, which typically proceeds via very elliptical orbits, i.e.
the stopping length is much greater than the impact parameter.

To recap, the dynamics of ultrarelativistic D-branes is strikingly
inelastic: copious production of highly-excited stretched open
strings rapidly drains the brane kinetic energy and traps the
branes into a tight orbit, eventually leading the branes to
coincide.

 In this simple and controllable example it proves possible to
 understand aspects of the backreaction
of open string production on the dynamics of colliding D-branes.
The lessons of our analysis could be extended to cosmological
models in which other sorts of fast-moving branes approach each
other and collide.  As we will discuss, these include the
ekpyrotic/cyclic universe scenario, brane-antibrane scenarios, and
the DBI model.

It is useful to indicate the various regions of parameter space
that we will probe.  We will outline this now to apprise the
reader of our strategy; later, in \S5.3, we will provide a more complete discussion.

 The dimensionless quantities of interest are the impact
 parameter $b$ measured in units of the string length; the string coupling
 $g_s$, which determines the mass of the D-branes in string units; and the
 initial relative velocity of the branes $v$.  We will find it more convenient to
convert this velocity into the rapidity, $ \eta \equiv
{\rm{arctanh}}(v)$.  We will usually set $\alpha' = 1$, except for
a few cases where we will retain explicit factors of the string
length for clarity.

Our goal in this work is to understand open string effects in {\it{relativistic}} dynamics;
the nonrelativistic case is already well-understood \cite{DKPS,KLLMMS}. We will therefore impose $1-v \ll 1$ so that $\eta \gg 1$.
Another important consideration is that the D-branes should have Compton wavelengths small compared to the impact parameter.
Because the D-branes grow light at strong string coupling, this amounts to a requirement that the coupling should be sufficiently weak.
Another obvious advantage of weak coupling is the suppression of string loop effects; our primary computation is a one-loop open string process.
 A further requirement is that the D-brane Schwarzschild radius should be much smaller than the impact parameter.  This too can be achieved with
a suitably small string coupling, as we will demonstrate in \S5.3.  Furthermore, although energy loss through closed string radiation can be
an important effect in a system of moving branes, there is a wide range
of string coupling, depending on $\eta$, for
which this effect is subleading compared to open string production.  Although all these considerations show that weak coupling is desirable
for control, it is important to recognize that as the coupling decreases, the D-branes grow heavy and hence stretch the strings
farther before coming to rest.

In summary, there is a range of values of the string coupling in which the backreaction of open strings is significant and competing effects
are suppressed.

The organization of this paper is as follows.  First, in \S2, we review the trapping of nonrelativistic branes, which provides
the basic intuition for the more complicated, stringy process which we aim to study.  Then, in \S3, we study the interaction
amplitude for moving branes.  We compute the brane interaction via an annulus diagram and examine its imaginary part, which
corresponds to open string pair production.  This result is well-known, but we include it for logical completeness and to set
our notation.  Our primary result appears in \S4, where we study the backreaction of open string production on the brane
trajectory and estimate the stopping length on energetic grounds.  In \S5 we discuss potential corrections and additional
effects, in particular the production of closed strings, and explain how they affect our considerations.  We conclude with a few
comments in \S6.  Finally, we collect useful identities about the
theta functions in the appendix.

\section{ Overview of the Trapping of Nonrelativistic Branes }

We will now briefly review the trapping of D-branes in
nonrelativistic motion, which was studied in \cite{KLLMMS}.  (See also \cite{Kofman, Felder} for earlier work on related mechanisms in field theory and cosmology.)  This
process is governed by pair production of massless open strings
and hence is describable in effective field theory.  It provides the basic framework
for understanding corrections to the brane dynamics, and so is a useful background for the
stringy trapping which we will study in \S3.

Because the field theory description is entirely sufficient, we can abstract the relevant properties of the worldvolume gauge
theory and represent the system with a simplified model,
\eqn{lagis}{ {\cal{L}} =
{1\over{2}}\partial_{\mu}\phi\partial^{\mu}\bar{\phi} +
{1\over{2}}\partial_{\mu}\chi\partial^{\mu}{\chi}-{g^2\over{2}}|\phi|^2\chi^2 }
in which a complex scalar field $\phi$ couples
to a real scalar field $\chi$.  At the origin $\phi=0$, $\chi$ becomes massless.

Let us consider the trajectory \eqn{trajis}{\phi(t) = i \mu + v t} in which $\phi$ is separated from the origin by the impact
parameter
$\mu$.  This is a solution to the classical equations of motion of \eno{lagis} provided that $\chi=0$.  Along this trajectory,
the mass of $\chi$ changes: in the limit where we impose \eno{trajis} and ignore the effect of the coupling to $\chi$, we may
rewrite \eno{lagis} as \eqn{newlag}{ {\cal{L}} = {1\over{2}}\partial_{\mu}\phi\partial^{\mu}\bar{\phi} +
{1\over{2}}\partial_{\mu}\chi\partial^{\mu}{\chi}-{g^2\over{2}}(\mu^2+v^2t^2)\chi^2 } so that the effective mass of $\chi$
varies with time.  This results in production of $\chi$ quanta.

This effect is easily understood in the quantum mechanics example of a harmonic oscillator whose frequency changes over
time from $\omega_{i}$ to $\omega_{f}$.  If the oscillator begins in its ground state at frequency $\omega_{i}$ but the
frequency changes nonadiabatically then the final state will not be the ground state of an oscillator of frequency
$\omega_{f}$.

One can readily compute the occupation numbers $n_{k}$ of modes with momentum $k$.  The result \cite{KLLMMS} is \eqn{bogis}{
n_{k} = \exp \Bigl(-\pi{{k^2+g^2\mu^2}\over{gv}} \Bigr).} If instead we consider a model in which the mass of $\chi$ is nowhere
zero, \eqn{masslagis}{ {\cal{L}} = {1\over{2}}\partial_{\mu}\phi\partial^{\mu}\bar{\phi} +
{1\over{2}}\partial_{\mu}\chi\partial^{\mu}{\chi}-\Bigl(m^2+{g^2\over{2}}|\phi|^2\Bigr)\chi^2 } the result is instead
\eqn{massbog}{n_{k} = \exp \Bigl(-\pi{{k^2+m^2+g^2\mu^2}\over{gv}} \Bigr).} The crucial, though intuitive, observation is that
production of a massive species is exponetially suppressed.  For this reason, production of massive string modes is entirely
negligible when the velocity is small.

We may now apply the result of the simplified model to a pair of D-branes.  Suppose that two Dp-branes, $p>0$, are arranged to
pass near each other.  The brane motion changes the masses
of stretched string states and induces pair production of unexcited stretched strings.  As the branes begin to separate, these
strings stretch and pull the branes back together.

This process can be followed in detail by numerically integrating the quantum-corrected equations of motion which follow from \eno{lagis}.
Such an analysis was presented in \cite{KLLMMS}.  However, analytical estimates are more readily generalized to the case of interest in this
paper, which is the stringy scattering of relativistic branes.  We will therefore explain how one can use energetics to estimate the stopping
length in the system \eno{lagis}.  (It was shown in \cite{KLLMMS} that such estimates are in excellent agreement with the numerical results,
although only the nonrelativistic case was studied there.)

After the branes have passed each other, the stretched open strings grow in mass.  Even though pair production has ceased, the energy contained
in open strings grows with time, because the strings are being stretched: \eqn{rhois}{ \rho_{open} \approx |\phi(t)|n_{open} }
When the energy in open strings is of the same order as the initial brane kinetic energy, the backreaction of the open strings is of order one and the brane motion slows
down significantly.  We therefore define the `stopping length' $\phi_{*}$
via $\rho_{open}(\phi_{*}) \approx {1\over{2 }}T_{p} v^2 $
where $T_{p}$ is the tension of a Dp-brane.

A few qualitative features of low-velocity trapping are worth mentioning.
First, the greater the number density of produced strings, the shorter the
stopping length.  On the other hand, the stopping length increases if the
brane velocity increases or the string coupling decreases (making the
branes heavier in string units).

The behavior in the limit $v \to 1$ is not obvious {\it{a
priori}}. To estimate the total number density $\nu_{total}$ of
all string modes, we could take the nonrelativistic result
\eno{massbog} for the occupation numbers of a massive species and
sum it over the levels $n$ in the string spectrum, including a factor of the density
of states $N(n)$.  The result (which was also presented in \cite{KLLMMS})
is \eqn{bogest}{ \nu_{total} \propto \sum_{n=0}^{\infty} N(n)
\exp\Bigl(-{2\pi^2\over{v}}\Bigl(n+{b^2\over{4\pi^2}}\Bigr)\Bigr).}
As we explain in \S4.1, the density of states at high levels $n$ obeys
\eqn{densis}{ N(n) \sim n^{-11/4} \exp\Bigl({\sqrt{8\pi^2 n}}\Bigr) }
This does not grow rapidly enough to compete with the exponential suppression \eno{massbog} of high levels, so
the limit $ v \to 1$ does not display strong production of excited strings.

However, we will show in detail in \S3, following \cite{Bachas}, that the actual number
density of produced strings is very much larger than the
nonrelativistic estimate \eno{bogest} suggests.  We will find
instead \eqn{betterest}{ \nu_{total} \propto \sum_{n=0}^{\infty}
N(n)
\exp\Bigl(-{2\pi^2\over{\eta}}\Bigl(n+{b^2\over{4\pi^2}}\Bigr)\Bigr)}
where $\eta = {\rm{arctanh}}(v) \gg 1$.  This result does not
follow from special relativity alone; it is instead a stringy
effect arising from velocity-dependent corrections to the
stretched string masses, as we will show.

\section{The Interaction Amplitude for Moving D-branes}

 We will now derive the interaction potential
 for two D-branes in relative motion with arbitrary velocity. Although this result is well-known
 \cite{Bachas}, we include the calculation for completeness
 and to set notation.

\subsection{Interaction Potential from the Annulus Diagram}

We will derive the interaction potential by computing the open
string one-loop vacuum energy diagram.  This diagram is an annulus
whose two boundaries correspond to the two D-branes. By the
optical theorem, twice the imaginary part of this amplitude is the
rate of pair production of on-shell open strings.  Thus, our goal
is to determine the imaginary part of the vacuum energy.

Several equivalent methods can be used to compute the vacuum
energy.  The original treatment \cite{Bachas} involves a direct
computation of the spectrum of open strings between the moving
branes; that is, it is possible to impose appropriate boundary
conditions and solve for the mode expansion. The vacuum energy is
then the sum of the zero-point energies of these oscillators.

We choose instead to review the perhaps more transparent computation given in \cite{JoeTwo}.
Let us stress that in this subsection we follow the treatment of \cite{JoeTwo} in detail, with very minor
modifications.

By double Wick rotation, a pair of branes in relative motion, separated by a transverse distance $b$, can be
mapped to a stationary pair of branes at an imaginary relative
angle, again separated by a distance $b$.  We will make this
precise below. Because the partition function for branes at angles
is very well understood, the vacuum energy is easily computed in
this approach.

Following \cite{JoeTwo}, we begin with two D4-branes which are parallel to each other, extended
along the directions $0,1,3,5,7$, and separated by a distance $b$ (the impact parameter) along $X^{9}$.
(To regulate the computation we compactify the spatial dimensions on a $T^{9}$ of radius $R$.)
Now let one brane move towards the other along the direction $X^{8}$ with velocity $v$.
That is, the coordinates of the moving brane are $X^{8} = v X^{0}, X^{9} = b$
while the other brane has $X^{8}=X^{9}=0$. This is our actual problem.

We now perform the Wick rotation $X^0 \to -i X'^7$, $X^7 \to  i X'^0$.
This transforms the moving branes into static branes which are misaligned by an
angle $\phi$ in the $(7',8)$ plane.  The angle $\phi$ is given by $X'^7 \tan \phi = X^8$.
The brane velocity $v$ and rapidity $\eta$ are related to this angle by
$\phi = - i~{\rm{arctanh}}(v) \equiv -i \eta $.

Next, it is useful to combine the coordinates into complex pairs
$Y_{a}$, where $Y_{1} = X^1 + i X^2, Y_{2}=X^3+i X^4, Y_{3} = X^5
+ i X^6, Y_{4} = X'^7 + i X^8$.  Define also the angles
$\phi_{1}=\phi_{2}=\phi_{3}=0,\phi_{4}=\phi$.  The rotation then
takes $Y_{4} \to \exp(i \phi) Y_{4}$.  It is now a simple matter
to set up the boundary conditions satisfied
 by strings which stretch between the branes:
 \eqn{Boundcond}{\eqalign{
& \sigma_1 = 0: \qquad \partial_1 \Re [Y_{a}] = \Im [Y_{a}] = 0 \cr
& \sigma_1 = \pi : \qquad \partial_1 \Re[\exp(i \phi_{a}) Y_{a}] = \Im[\exp(i
 \phi_{a}) Y_{a}]=0 \,.
}}
The solutions to the wave equation which satisfy these boundary
 conditions are:
 \eqn{Modes}{
 Y_{a}(w, {\bar w}) = \exp(-2i \phi_{a}) \left[ i {\sqrt {\alpha' \over 2}}
 \sum_{r = {\bf Z} + \phi_{a}/\pi} {\alpha_r^a \over r} \exp(irw) \right]
 + \left[- i {\sqrt {\alpha' \over 2}}
 \sum_{r = {\bf Z} + \phi_{a}/\pi} {\alpha_r^a \over r}^* \exp(ir{\bar w})
 \right] \,,
}
where $w = \sigma_1 + i \sigma_2$. We can readily write down the
 partition function for these four scalars:

%\eqn{ScalarZ}{
% Z_{scalar}(\phi_{a}) = q^{E_0} \prod_{m=0}^{\infty} \left[1-q^{m + (\phi_{a}/\pi)} \right]^{-1}
% \left[1-q^{m + 1 - (\phi_{a}/\pi)} \right]^{-1} = - i {{\exp(\phi_{a}^2
% t/\pi) \eta(it)} \over {\theta_{11} (i \phi_{a} t/\pi, it)}} \,,
%}
%where $q = \exp(-2 \pi t).

\eqn{ScalarZ}{Z_{scalar}(\phi_{a}) =  - i {{\exp(\phi_{a}^2
t/\pi) \eta(it)} \over {\theta_{11} (i \phi_{a} t/\pi, it)}}}
so that the resulting bosonic partition function is
\eqn{zbos}{ Z_{boson} = \prod_{a=1}^{4} Z_{scalar}(\phi_{a}) }
In a similar way, one can compute the fermionic partition
 function, keeping in mind the various spin structures:
\eqn{FermZ}{
 Z_{ferm} = \prod_{a=1}^{4}Z^1_1({\phi_{a}/2},it) \,,
}
where
\eqn{MoreDef}{ Z_1^1({\phi_{a}/2},it) \equiv  {\theta_{11}(i \phi_{a} t/{2\pi},
  it) \over {\rm{exp}}(\phi_{a}^{2} t/{4\pi}) \eta(it)}}

%\eqn{MoreDef}{\eqalign{
%& Z_1^1({\phi/2},it) \equiv  {\theta_{11}(i {\phi/2} t/\pi,
%  it) \over {\rm{exp}(\phi^2 t/\pi) \eta(it)}}  \qquad a = 1,2,3,4 \cr
%& \phi'_1 = {1 \over 2}(\phi_1 + \phi_2 + \phi_3 + \phi_4) \qquad
%  \phi'_2 = {1 \over 2}(\phi_1 + \phi_2 - \phi_3 - \phi_4) \cr
%& \phi'_3 = {1 \over 2}(\phi_1 - \phi_2 + \phi_3 - \phi_4) \qquad
%  \phi'_4 = {1 \over 2}(\phi_1 - \phi_2 - \phi_3 + \phi_4) \,.
%}}

We conclude that the one-loop potential is
\eqn{CrudeLoopAngle}{
 V = - \int_0^{\infty} {dt \over t} {1 \over {\sqrt {8 \pi^2 \alpha'
       t}}} \exp \left(-{{t b^2} \over {2 \pi \alpha'}}\right)
 \prod_{a=1}^4 {\theta_{11}(i \phi_a t/{2\pi}, it) \over \theta_{11}(i
   \phi_a t/\pi, it)} \,.
}

This potential governs D4-branes at a relative angle.  To map into the case of interest,
we T-dualize as many times as needed, each time introducing the replacement  $\theta_{11}(i \phi_a
t/\pi, it) \to  i {\sqrt{8 \pi^2 \alpha' t} \eta^3(it)}/R,$ where $R$ is the size of the spatial torus.

This finally brings us to the potential for
p-branes at an angle $\phi$:
\eqn{LoopAngle}{
 V = - i R^{p} \int_0^{\infty} {dt \over t} (8 \pi^2 \alpha' t)^{-p/2} \exp \left(-{{t b^2} \over {2 \pi \alpha'}}\right)
  {\theta_{11}(i \phi t/2\pi, it)^4 \over {\theta_{11}(i
   \phi t/\pi, it) \eta(it)^9}} \,.
}
Our final interest is in the number density and energy density of
open strings, so the spatial volume $R^{p} \equiv i V_{p} $ will eventually cancel.

To read off the desired result for moving branes, we set $\phi = - i \eta$ to get
\eqn{LoopMov}{
 V = V_p \int_0^{\infty} {dt \over t} (8 \pi^2 \alpha' t)^{-p/2}
 \exp \left(-{{t b^2} \over {2 \pi \alpha'}}\right)
 {\theta_{11}(\eta t/2 \pi, it)^4 \over {\theta_{11}(
   \eta t/\pi, it)\eta(it)^{9}}} \,.
}

One can easily show that this agrees precisely with the result of
\cite{Bachas}, equation (11).  To see this, use \eno{Eta} and
\eno{secondident}, define $t_{there}= 2 t$, $ \epsilon =
{\eta\over{\pi}}$, and set $\alpha'={1\over{2}}$.

A useful equivalent form for \eno{LoopMov} is

\eqn{tauloop}{ V = V_{p} \int_{-\infty}^{\infty} d \tau
\int_0^{\infty} {dt \over t} (8 \pi^2 \alpha'
t)^{-p/2}{\theta_{11}(\eta t/2 \pi, it)^4 \over {\theta_{11}(\eta
t/\pi, it)\eta(it)^{9}}}\exp \left(-{{t\over {2 \pi
\alpha'}}\left(b^2+v^2\tau^2\right)}\right) \times  {v
\over \pi} \sqrt{{t\over{2\alpha'}}} } In this form the time-dependence of
the stretched string masses is manifest.

\subsection{Imaginary Part and Pair-Production Rate}

The above expression from the interaction potential is rich in
information. The real part tells us about the velocity-dependent forces
from closed string exchange, while twice the imaginary part is equal
to the rate of production of open strings.

The potential \eno{LoopMov} would be real if the integrand had no poles.
However, $\theta_{11}(\eta t/\pi,it)$ has a zero for integral values of $\eta t/\pi
\equiv k$, so we can compute the imaginary part of the integral by summing the
residues at the corresponding poles.

\eqn{PairProd}{
 \Im[V] = {V_p \over {2 (2 \pi)^p}} \sum_{k=1}^{\infty} {1 \over k}
 \left(\eta \over{\pi k} \right)^{p/2} \exp\Bigl(-{{b^2 k}\over{2 \eta}}\Bigr)
 Z(i k\pi/\eta)\Bigl(1-(-1)^{k}\Bigr) \,, }
where we have defined the partition function $Z(\tau) \equiv
{1\over{2}}\theta^{4}_{10}(0|\tau)\eta(\tau)^{-12}.$
(The factor projecting out even values of $k$ arises because of Jacobi's `abstruse identity'.)

This expression, which was first derived in \cite{Bachas}, will be essential to our investigation.  By
extracting its behavior in various limits we will be able to study the
effect of open string production on the brane dynamics.

First of all, we can check the normalization of \eno{PairProd} by taking
the low-velocity limit, in which $\eta \to v$.  The result is
\eqn{Schwinger}{
 \Im[V] = {{8 V_p} \over {(2 \pi)^p}} \sum_{k=1,3,5, \ldots}^{\infty}
 {1 \over k} \left(v \over {\pi k} \right)^{p/2} \exp\Bigl(-{b^2 k\over{2 v}}\Bigr) \,. } This
is identical to Schwinger's classic result \cite{Schwinger} for the
pair-production rate of electrons in a constant electric field.  In the
present case, the interpretation is of pair production of massless open
strings between the branes, which was also obtained by the method of
Bogoliubov coefficients in \cite{KLLMMS}.

Our interest is in the case of velocities approaching the speed of light.
We expect that the dominant contribution to pair production in this limit
will come from highly-excited string states.  Because the density of
states grows exponentially \eno{densis} at high levels, we
anticipate copious production of massive strings and, as a result,
dramatic backreaction on the brane motion.

To investigate this, we begin with the high-velocity limit $\eta \gg 1$ of
\eno{PairProd}:  \eqn{ultraseries}{ \Im[V] = {V_{p}\over{2(2\pi)^p
}}\sum_{k=1,3,5,\ldots}^{\infty}
{1\over{k}}\left({\eta\over{\pi k}}\right)^{p/2-4} \times
\exp\Bigl({\eta\over{k}} -{b^2 k\over{{2 \eta}}}\Bigr)\Bigl(1+{\cal{O}}(e^{-{
\eta/k}})\Bigr)} where we have used the asymptotics \eno{asympt}.

Keeping the dominant contribution, which comes from $k=1$,
and expressing the result as a number density $\nu_{open}$
of open strings stretching between the branes, we find \eqn{FastProd}{
 \nu_{open} \approx c_{p} \eta^{{p\over{2}}
-4} \times \exp\Bigl(\eta - {b^2\over{2\eta}}\Bigr) \,
}
where $c_{p} = \Bigl({2 (2 \pi)^p \pi^{p/2 -4}}\Bigr)^{-1}.$

There are three important differences between the low-velocity effect
in \cite{Schwinger} and the high-velocity relation of \eno{FastProd} . The
first is that production of strings is exponentially suppressed at low
velocities: this can be understood from the fact that the amount of
strings produced at a given energy falls off exponentially with energy,
while the density of states for such low energies is a simple power law.
At high energies, however, the density of states grows exponentially and
these two competing exponentials lead to copious string production if the
initial velocity of the branes is sufficiently high.

The second important difference is that at low velocities, the efficacy of the trapping process is strongly dependent on the impact
parameter. For large impact parameters, $b \gg 1$, (recall that $b$ is measured in string units) the trapping is exponentially weak.
For ultrarelativistic branes, however, the trapping weakens only when $b \gg \eta$.  The effective range of strong trapping is evidently
much increased in the ultrarelativistic limit.

Finally, in the low-velocity limit, the energy of produced open strings is
a negligible fraction of the D-brane energy \cite{KLLMMS} until the branes
separate far enough to stretch the open strings significantly.  The
associated distance, the `stopping length', is generically much larger than the impact parameter.
In the ultrarelativistic limit, in contrast, the energy
carried by the open strings can be comparable to the brane kinetic energy
even before any stretching.  This occurs because high speeds make possible
the production of highly-excited strings with significant oscillator
energy.  This consideration suggests that the backreaction of open strings
is much more dramatic for relativistic branes than for nonrelativistic
ones.  We undertake a careful study of this in the following section.

\section{ Backreaction from Energetics}

We have seen in the previous section that relativistic brane
motion leads to the production of a tremendous number density
\eno{FastProd} of stretched open strings.  We would now like to
estimate the effect of this process on the brane motion, and to do
so we must estimate the energy density carried by the produced
open strings.

\subsection{Open String Energy}

This energy is easily computed if we first rewrite the partition function
$Z$ as a sum over string states.  This is conveniently parametrized in terms of the excitation level $n$
and the number of states $N(n)$ at each level.
\eqn{partition}{ Z(i k \pi/\eta) \equiv
{1\over{2}}\theta_{10}(0,ik\pi/\eta)^4\eta(ik\pi/\eta)^{-12} =
\sum_{n=0}^{\infty} N(n) \exp\Bigl(-{2\pi^2 n k\over{\eta}}\Bigr).}
We would first like to determine the behavior of $N(n)$ at high excitation levels $n$.  Taking the ansatz
\eqn{nansatz}{ N(n) \approx c_{N} n^{a} \exp\Bigl(b \sqrt{n}\Bigr),}
approximating the sum by an integral, evaluating this integral by stationary phase, and demanding the asymptotics \eno{asympt},
we find \eqn{nisnow}{ N(n) \approx {(2n)}^{-11/4} \exp\Bigl(\pi\sqrt{8 n} \Bigr).}
The numerical prefactor was chosen for convenience; strictly speaking, the approximate evaluation
of the integral does not determine
constant prefactors of order unity, but for our purposes it suffices to choose the factor now as in \eno{nisnow}.

With this result in hand, we can rewrite \eno{PairProd} as
\eqn{summingspectrum}{
 \Im[V] = {V_p \over {(2 \pi)^p}} \sum_{k=1,3,\ldots}^{\infty} {1 \over k}
 \left(\eta \over {\pi k} \right)^{p/2} \exp\Bigl(-{b^2 k\over{2 \eta}}\Bigr)
  \sum_{n=0}^{\infty} N(n) \exp\Bigl(-{2\pi^2 n k\over{\eta}}\Bigr).}

An equivalent form for this relation is
\eqn{tauPairProd}{  \Im[V] =  {{ \sqrt{2} v V_p }\over {(2 \pi)^{p+1}}}
\int_{-\infty}^{\infty} d\tau \sum_{k=1,3,\ldots}^{\infty} {1 \over k}
 \left(\eta \over {\pi k} \right)^{{p-1\over{2}}}
  \sum_{n=0}^{\infty} N(n) \exp\Bigl(-{{2 \pi^2 k M^2(\tau) }\over{\eta}}\Bigr).}
where
\eqn{massis}{ M^2(\tau) \equiv {n\over{\alpha'}} + {{b^2+v^2\tau^2}\over{4\pi^2\alpha'^2}}.}
To determine the total energy of the produced strings, we begin with the energy of a string at level $n$, when the
separation of the branes along the direction of motion is $r$:
\eqn{stateenergy}{E(n)^2 = {v^2\over{\eta^2}}\Bigl( {n\over{\alpha'}}+ {{b^2 + r^2}\over{4\pi^2\alpha'^2}}\Bigr) }
The velocity-dependence is perhaps counterintuitive, but can be derived by requiring consistency
of the annulus result \eno{PairProd} with a steepest-descent computation in the nearly-adiabatic
limit $b \gg \eta, \eta \gg 1$.  However, we are not aware of a simple, {\it{a priori}}
computation of this mass correction.  Note that the mass \eno{stateenergy} does reduce to the
correct rest mass in the limit of small velocity.

We can now express the energy density of produced
open strings as
\eqn{summingenergy}{
 \rho_{open} = {1\over{(2\pi)^p}} \sum_{k=1,3,\ldots}^{\infty} {1
\over k}
 \left(\eta \over {\pi k} \right)^{p/2} \exp\Bigl(-{b^2 k\over{2\eta}}\Bigr)
  \sum_{n=0}^{\infty} E(n) N(n) \exp\Bigl(-{2\pi^2 n k\over{\eta}}\Bigr)
\,. }

Because of the competition of the growing and decaying exponential factors, this sum is dominated by terms near some $n_{peak} \gg 1.$
As indicated above, we approximate the sum on levels using the relation
\eqn{sumrep}{ \sum_{n=0}^{\infty} N(n) n^{\alpha} \exp\Bigl(-{2\pi^2 n \over{\eta}} \Bigr) \approx 2^{-11/4} \int_{n_{0}}^{\infty} dn n^{\alpha -11/4}  \exp\Bigl(\pi\sqrt{8 n} - {2\pi^2 n \over{\eta}} \Bigr)}
where the lower bound $n_{0} >0$ is chosen so that the integral is dominated by $n\approx n_{peak}$, not $n \approx 0$.  We have kept the leading term in the sum on $k.$
By the method of stationary phase we find that the integral is dominated by $n \approx n_{peak} = {\eta^2}{(2\pi^2)}^{-1},$ leading to
\eqn{statphase}{ 2^{-11/4} \int_{n_{0}}^{\infty} dn n^{\alpha-{11/4}} \exp\Bigl(\pi\sqrt{8 n} - {2\pi^2 n \over{\eta}} \Bigr) \approx {1\over{2}}e^{\eta}\Bigl({\pi\over{\eta}}\Bigr)^{4} \Bigl({\eta^2\over{2\pi^2}}\Bigr)^{\alpha}.}
For $\alpha=0$ this reproduces the asymptotic behavior \eno{asympt}; we normalized \eno{nisnow} to arrange this.

This approximate result provides an important physical lesson: the primary contribution to the open string energy comes from strings at levels $2\pi^2 n \approx \eta^2$.
For such a string, \eqn{eisnow}{ E(n) = {v\over{2\pi\eta}} \sqrt{ 4\pi^2 n + b^2 + r^2 } \approx
{1\over{2\pi}}\sqrt{ 2 + {{b^2 + r^2}\over{\eta^2}} } .}

Let us now examine this result in the parameter ranges of interest.  If the stretched string length is large compared to $\eta$, $\sqrt{b^2+r^2} \gg \eta$, then the sum \eno{summingenergy}
is simply \eqn{esimp}{ \rho_{open} \approx {\sqrt{b^2+r^2}\over{2 \pi \eta }}\nu_{open}.}  On the
other hand, when the rapidity is larger than the separation,
$\eta \gg \sqrt{b^2+r^2}$, we have instead \eqn{eeta}{ \rho_{open} \approx
{1\over{\pi\sqrt{2}}}\nu_{open}} where we have used \eno{statphase} with $\alpha = {1/2}.$

The key observation which follows from \eno{eeta} is that the
energy density carried by produced pairs of stretched open
strings can be a significant fraction of the kinetic energy
density of the Dp-brane.  The backreaction from open string production is
therefore an important contribution to the dynamics of relativistic
D-branes.  We will now examine this in detail.

\subsection{Estimate of the Stopping Length}

It will be very important to recognize three length-scales which arise in
the problem: the effective size $r_{eff}(\eta)$ of a relativistic brane, the critical impact parameter $b_{crit}(\eta)$ beyond which the trapping rapidly weakens,
and the size $r_{nad}(\eta)$ of the region in which the stretched open string masses change nonadiabatically.

The factor depending on $b$ in \eno{FastProd} indicates that the effective area of a brane moving with rapidity $\eta$
is \cite{Bachas} \eqn{reffis}{ r_{eff}^2 \approx \eta \alpha'.}  This corresponds precisely to
the logarithmic growth in cross-sectional area of a highly-boosted
fundamental string, $ r_{eff}^2 \sim \alpha'~{\rm{ln}}(\alpha' s)$, where
$\sqrt{s}$ is the center-of-mass energy.  The explanation for this growth
is that a Regge probe of an ultrarelativistic string is sensitive to
rather high-frequency virtual strings, whose considerable length creates a
large cloud of virtual strings \cite{JoeOne}.  We conclude that a D-brane
with rapidity $\eta$ has an apparent radius $r_{eff} = \sqrt{\eta \alpha'}.$

The growth in effective area provides an additional perspective on the
velocity-dependent mass \eno{stateenergy} of stretched strings.  Both
effects may be considered to originate in a rescaling of the effective
string tension, \eqn{teta}{ T(\eta) = {1\over{2\pi\alpha'\eta}}}
This results in a D-brane
size $r_{eff} \approx \sqrt{\eta \alpha'}$ and a stretched string mass
\eqn{effmass}{ m_{eff} = {1\over{\sqrt{\eta\alpha'}}}\sqrt{n+
{b^2\over{4\pi^2}}}} consistent with \eno{reffis},\eno{stateenergy}.  This
rescaling of the effective tension is a very useful heuristic for
understanding the dramatic difference between the naive result \eno{bogest}
and the complete annulus computation \eno{FastProd} for the number density.
This rescaling can also be understood from the T-dual electric field
perspective: as the electric field approaches a critical value, the strings
can no longer hold themselves together, so their effective tension goes to
zero \cite{Porrati}.

Next, to find the critical impact parameter, we note that the open string
energy density obeys
\eqn{reminder}{\rho_{open}
\propto \exp\Bigl(\eta -
{b^2\over{2\eta}}\Bigr),} so that for $\eta \gg 1,$ the critical distance is evidently
$b_{crit}\sim \eta.$ For impact parameters less than $b_{crit}$, the open
string energy
density is generically large. The trapping effect is therefore very
strong for impact parameters of order $b_{crit}$ and
smaller. (Nevertheless, trapping still occurs for impact parameters
much larger than $b_{crit}$.)

%From \eno{statphase} one can see that the open string energy
%at minimum brane separation is primarily the oscillator energy of strings at level $n \approx {\eta^2\over{2\pi^2}}.$

Finally, the nonadiabaticity is characterized by how rapidly the frequency changes
with time.  Quantitatively, it is measured by the dimensionless quantity
$\xi \equiv {\dot{\omega}\over{\omega^2}},$ where $\omega$ is the frequency.  Using \eno{massis} we find
\eqn{nadis}{ \xi = {{2\pi \eta r \dot{r}} \over{ (4\pi^2 n + b^2 +
r^2)^{3/2} }} \approx {{2\pi r \eta } \over{( 2\eta^2 + b^2 + r^2)^{3/2}
}}}
which reaches its peak at $ r^2 = \eta^2 + {1\over{2}}b^2.$  Thus, the region of nonadiabaticity has size $r_{nad} \sim \eta$.
Open strings are produced in large quantities when $ - r_{nad} \lsim r \lsim r_{nad}.$

In summary, the critical impact parameter is $b_{crit} \sim \eta,$ which is also the size of the nonadiabatic region.
 The effective radius of a moving D-brane, i.e. the size of the stringy halo, is much smaller,
 $r_{eff} \sim \sqrt{\eta} \ll b_{crit}$.  For any fixed, large $\eta$ we can require
\eqn{paramis}{ r_{eff} \ll b \ll b_{crit}} so that the trapping is very strong but the stringy halos
 are small enough to be unimportant.  The case of a head-on collision, $ b \lsim r_{eff},$ is also interesting,
particularly for the question of string production in the cyclic universe models, but we will first
explore the better-controlled regime \eno{paramis}.

One further observation is that a scattering event with impact parameter $b$, no matter how
powerful the trapping, will typically involve motion on an arc
whose initial radius is at least of order $b$.  Angular momentum
conservation prevents the moving brane from coming abruptly to a
complete stop; over one or more orbits, however, there is
sufficient time to radiate away the angular momentum into closed
string modes, as we will see in \S5.1.  In strong trapping, such
as we will find in the relativistic case, the orbits will be
roughly circular, whereas in nonrelativistic trapping the typical
orbit is very highly elliptical, indicating weaker binding.

With these estimates in hand we can at last compute the stopping length for a scattering event.
Taking one brane to be at rest and the other to have velocity $v$,
we define as before \eqn{etais}{\eta \equiv {\rm{arctanh}}(v).}
Working instead in the center of mass frame, the branes
approach each other with velocities \eqn{upsilonis}{ u =
\tanh(\omega) = \tanh(\eta/2) } so that the center-of-mass
$\gamma$ factor for either brane is \eqn{gammais}{ \gamma =
{1\over{\sqrt{1-u^2}}} \sim {1\over{2}}e^{\omega}} when $\omega
\gg 1$. The energy density of the brane pair is then \eqn{epair}{
E_{tot}=2 T_{Dp} \gamma \sim T_{Dp} e^{\omega} = T_{Dp}
e^{\eta/2}.}

We therefore find that for $\eta \to \infty$,
\eqn{energetics}{\rho_{Dp} = T_{Dp} e^{\eta/2}
  = {1 \over {g_s (2 \pi)^p}} e^{\eta/2} .}
In the case of strong trapping, $ b \ll b_{crit} \approx \eta$, the open string energy
at the minimum brane separation is \eqn{open}{\rho_{open} \approx {c_p\over{\pi\sqrt{2}}}
{\eta^{{p\over{2}}-4} \exp \Bigl({\eta}\Bigr)}}
whereas for weak trapping, $ b \gg b_{crit}$, the open string energy is instead
\eqn{weakopen}{\rho_{open} \approx {c_p\over{2\pi}} {\eta^{{p\over{2}}-5} \sqrt{b^2 +
      r^2} \exp \Bigl(\eta - {b^2\over{2\eta}} \Bigr)} ,} where $c_p = \Bigl({2 (2 \pi)^p \pi^{p/2
-4}}\Bigr)^{-1}.$
Of course, the open string energy depends on $r$ even in the case of strong trapping, but this dependence is relatively unimportant until $r \sim \eta.$

Comparing \eno{energetics},\eno{open} we conclude that if an
external force compels the branes to pass each other at constant,
ultrarelativistic velocity, then, unless the string coupling is
exponentially small, the energy stored in open strings at the
point of closest approach is considerably larger than the initial
kinetic energy of the branes. This means that without an
artificial external force, the branes will {\it{not}} pass each other
with undiminished speed, as this is energetically inconsistent.

We expect instead that as open strings are produced, the branes
slow down gradually, leading to diminished further production of
strings.  The final result, of course, will be consistent with
conservation of energy.  (In \S4.3 we will address the production
of open strings between decelerating branes, and in \S5.1 we will
explain that the emission of closed string radiation also serves
to reduce the rate of production of open strings.)

Although the open string energy in \eno{open} is an overestimate for
the reason just mentioned, we will nevertheless use it now to find an
estimate of the stopping length.  This will serve to illustrate our
technique in a manageable setting; it will then be a simple matter to
repeat the analysis including the corrections of \S4.3 and \S5.1, which will not alter the form of our result.

We define the stopping length $r_{*}$ by $\rho_{open}(r_{*}) = \rho_{Dp},$ so that at $r=r_{*}$
all the initial energy has been stored in stretched open strings.
Equating \eno{energetics} and \eno{weakopen}, we find the stopping length
\eqn{stopis}{
r_{*}
\approx {4 \pi^2 \over{g_{s}}}
\exp\Bigl(-{\eta\over{2}}+{b^2\over{2\eta}}\Bigr)\left({\eta\over{\pi}}\right)^{5-p/2}.}
This is our main result.  It manifests the surprising property that for sufficiently large rapidity, the stopping length $\it{decreases}$ as the rapidity increases.
(More precisely, for any fixed $g_s,b$ there exists a rapidity $\eta_{min}$ such that the stopping length decreases as $\eta$ increases past $\eta_{min}$.)  To
 understand this unusual property, it is useful to keep in mind the behavior of D-branes scattering at even greater speeds, so great that the stringy halos themselves collide.
For any $b$ there is an $\eta$ such that $r_{eff} \gsim b$; the scattering of the branes is then described by the collision of
 absorptive disks of radius $r_{eff}$ \cite{Bachas}.  Moreover, for a suitable range of $g_s$ the brane Schwarzschild radii are so large that black hole production is an important consideration.  We have carefully chosen our parameter ranges to exclude these effects and focus instead on the more controllable regime of strong stringy trapping; however, the black disk collisions and black hole production serve to illustrate that the limit of arbitrarily high rapidity involves very hard scattering and high inelasticity, in good agreement with the large-$\eta$ behavior of \eno{stopis}.

The stopping length \eno{stopis} is large in string units only when \eqn{gsis}{ g_{s} \ll~
4\pi^2
\exp\Bigl(-{\eta\over{2}}+{b^2\over{2\eta}}\Bigr)\left({\eta\over\pi}\right)^{5-p/2}} which
is
an exponentially small value of the coupling provided $\eta \gg b, \eta \gg 1.$ Thus, although
backreaction from open string production is a higher-order correction to
the dynamics \cite{BachasLectures} which one might suppose is unimportant
at moderately weak coupling, we have shown that for relativistic branes with $ b \ll \eta$
the backreaction of open strings is crucial unless the string coupling is extraordinarily small.

\subsection{Corrections from Deceleration }

All of our computations so far have applied exclusively to a pair
of branes approaching each other at constant velocity. On the
other hand, we have demonstrated that the backreaction from open
string production, as computed along this trajectory, necessarily
causes the branes to decelerate. Clearly, the next step is to
understand how the amount of string production changes when the
branes follow a decelerating trajectory.

The analysis of string production during deceleration turns out to
be tractable in the nonrelativistic limit.  However, we have not
found an exact answer for the relativistic case.  Upon double Wick
rotation the amount of string production between decelerating
branes is mapped to the interaction between curved branes, which
is not obviously solvable with conformal field theory techniques.

Even though we will not find an exact result for the string
production, we will be able to place bounds on the resulting
number density.  This is sufficient information for the analysis
of \S4.3: although our results there for the stopping length will
not be as precise as they are in the nonrelativistic limit, the
qualitative features -- copious production of excited strings,
rapid trapping, and very high inelasticity -- will be quite clear.

First, however, we will examine the limit of instantaneous
deceleration. Take the branes to move with a velocity $v_0$ for all
$t<0$, but to come to rest for
 $t>0$. This problem can be solved exactly by matching the
parabolic cylinder functions (and their derivatives) to the plane
wave solutions at $t=0$. However, this setup clearly involves
enormous non-adiabaticity and so there would be an extremely large
amount of pair-production, far greater even than in the case of
constant velocity.  This is readily computed, but it is not
useful; we would like a more conservative estimate.

A more realistic picture is one in which the relative velocity of
the branes varies as a function of time, for example as $v(t) =
v_0 (1 -
 \tanh(t/f))$, where $f$ measures how abruptly the brane slows down.
(Note also that in this setup the initial velocity is $v(-\infty) = 2v_0$.)
The wave equation governing the stretched strings is therefore
\eqn{Decel}{
\Bigl(\partial_t^2 + k^2 + g^2 b^2 + g^2 v_{0}^2 [t - \log(\cosh (t/f))]^2
\Bigr) \chi = 0 \,.
}

It is instructive to consider the non-adiabaticity parameter $\xi \equiv {\dot \omega}
/\omega^2,$ where \eqn{omeis}{\omega^2(t) = k^2 + g^2 b^2 + g^2 v_{0}^2 [t -
\log(\cosh (t/f))]^2.}  Let us first take $f \ll 1,$ which is the case of very
rapid deceleration.  In this limit the deceleration is concentrated at $t=0$,
so that for slightly later times, when the branes have come to a
halt, we have $\xi =0$ and hence no
particle production.  Comparing this scenario to that of branes moving with uniform velocity $2v$ and no deceleration,
we see that an abrupt stop reduces the effective time available for particle production by a factor of two.  Thus, for branes which come to a halt very rapidly,
the total number of particles produced is approximately half the number produced when the branes move with uniform velocity.

We can analytically solve the problem in the opposite limit of very gentle deceleration, $ f \gg
{\sqrt{k^2 + g^2 b^2}}/(g v_{0}).$   Using the
steepest descent method to determine the Bogoliubov coefficients
\cite{Chung, Gubser, GubRev} and observing that in this limit there is a branch point very near
the imaginary axis, at $ - i {\sqrt {k^2 + g^2 b^2}}/(g v_{0}),$ we find \eqn{steepbeta}{|\beta_{k}|^2 = \exp\Bigl(- \pi (k^2 + g^2 b^2)/(g
v_{0})\Bigr).}  This coincides with the exact result for the
constant-velocity problem with velocity $v(t)= v_{0}$.  However, as we already noted, in the present case
the initial velocity is $v(-\infty) = 2 v_{0}$.  Our very simple conclusion is that this gradually decelerating trajectory leads to the
same amount of string production as an unaccelerated trajectory in which the branes move at a uniform velocity which is smaller by a factor of two.
The effective velocity, for purposes of particle production, is thus the average velocity ${1\over{2}}(v(-\infty)+v(\infty))$.

We conclude that very gradual deceleration results in significantly reduced string production.  In particular, comparing the limits of
large and small $f$, we see that the reduction in number density is much greater for gradual than for rapid deceleration.

The above result applies to nonrelativistic motion.  The string
computation which would be analogous to the annulus partition
function but incorporate deceleration is considerably more
complicated.  In particular, the acceleration of the branes breaks
conformal invariance, so it is difficult to use conventional
techniques to compute the string production in this case.

Fortunately, it is possible to estimate the stopping length
without an exact result for the string production during
deceleration.  The simple argument relies only on energetics and
on the constant-velocity result \eno{FastProd}.

Suppose that open string production slows a moving brane, bringing
it from an initial kinetic energy $E_{i} = \gamma_{i} T_{p}$ to an energy
(at the point of closest approach) $E_{f} = \gamma_{f} T_{p}$, where
$\gamma_{i},\gamma_{f}$ are the usual relativistic factors.  The
stopping length, defined again by $E_{i} = E_{open}(r_{*})$, is easily
seen to be \eqn{decelstop}{
{r_{*}} \approx  {2\pi\eta E_{i}\over{\nu_{open}}} =
{{\sqrt{2}\eta}}{E_{i}\over{E_{i}-E_{f}}},} where we have used
\eno{esimp},\eno{eeta}.

Consider first the case $\gamma_{f}
\gg 1$.  If the stopping length is large compared to the size $r_{nad}$ of
the nonadiabatic region, $r_{*} \gg \eta $, then the branes are moving quickly as they leave the region of nonadiabaticity.
This means that the result \eno{FastProd} applies directly, and we return to an
apparent inconsistency: the open string energy is large compared
to the initial energy.  This is a clear signal that the stopping
length cannot be much larger than $r_{nad} \sim \eta $.

A stopping length of order $\eta $ or smaller is indicative of strong trapping: the branes come to rest
around the time that the nonadiabaticity grows small, which means that a few strings are still being produced.
Furthermore, this distance $\eta$ is comparable to the critical impact parameter and critical orbital radius.

On the other hand, in the case $\gamma_{f} \sim 1$, we have $E_{f}
\ll E_{i}$, so that \eno{decelstop} yields the stopping length $r_{*}
\approx \sqrt{2}\eta$.

We conclude that no matter how the deceleration affects open
string production, if the only process acting to slow the branes
is loss of energy to open strings, then the stopping length is no
more than of order $\eta $, i.e. the size of the nonadiabatic
region.  Thus, the trapping is very strong: very little stretching is required before the branes are brought to rest.

Given a good estimate of the open string production along a
decelerating path, we could give a more accurate estimate of the
stopping length.  However, we have just demonstrated through
energetics and the result \eno{FastProd} that in any event this
stopping length is no larger than $\eta$.  In fact, we
expect that it is actually considerably smaller than this.

It remains a possibility that loss of energy through closed string
radiation could modify this result.  We now proceed to show that
this is not the case.

\section{ Further Considerations}

\subsection{ Production of Closed Strings }

By incorporating the effects of open string production we have
seen that relativistic D-branes decelerate abruptly as they pass
each other.  This deceleration will lead to radiation of closed
strings, in a process analogous to bremsstrahlung.  This drains
energy from the brane motion, and, unlike the transfer of energy
into stretched open strings, this energy is forever lost from the
brane system.  Closed string radiation therefore serves to
increase the inelasticity of a brane collision.  Now, the end
state of a near-miss is a spinning `remnant', i.e. two D-branes
orbiting rapidly around each other, connected by a high density of
strings.  Loss of energy and angular momentum to closed string
radiation will swiftly reduce the rotation of this remnant, at
least until the velocities become nonrelativistic.

One potential worry is that the energy loss to radiation might be
so large that the quantity of open strings produced during a
near-miss is quite small, leading to weak trapping and a large
stopping length.  This is an example of the more general concern
that string production could be highly suppressed if any other
effect caused the branes to decelerate to nonrelativistic speeds
before reaching each other.  We will show that the radiation of
massless closed strings can be energetically significant but, even
so, does not alter our conclusion that the stopping length is not
large in string units.

To estimate the energy emitted as massless closed strings, we will
make use of the close analogy of this process to gravitational
bremsstrahlung \cite{Peters} and to gravitational synchrotron
radiation \cite{Breuer}.  Of course, one of the massless closed
string modes is the graviton, but we also expect radiation of
scalars, including the dilaton and, when present, the
compactification moduli.  Even so, it will not be at all difficult
to convert results from general relativity to the case at hand,
because in practice, relativists often use the far simpler scalar
radiation to estimate the basic properties of gravitational
radiation.  We will do the same.

Consider a small mass $m$ moving rapidly past a large mass $M$ in
a path which is, to first approximation, a straight line.  A burst
of gravitational radiation will be emitted in a very short time,
at the moment of closest approach.  This is called gravitational
bremsstrahlung.  The peak radiated power is approximately
\cite{Peters} \eqn{bremis}{ P \sim {G^{3} M^2 m^2 \over{ b^4}}
\gamma^4} where $G$ is the Newton constant, $b$ is the impact
parameter, and $\gamma$ is the relativistic factor.  For the
remainder of this section we omit numerical prefactors: it will
suffice to have the dimensional factors and the powers of
$\gamma$.

The case of interest to us is extremely strong binding by open
strings, for if the acceleration caused by the open strings is
small then the closed string radiation should not play a key role,
and the argument for trapping given in \S4.3 suffices.  Thus, we
model the brane scattering by a gravitational scattering event in
which the impact parameter is not much larger than the
Schwarzschild radius of the larger mass.  This gives
\eqn{bremisnow}{ P \sim {G m^2\over{b^2}} \gamma^4.}

Another useful case is that of gravitational synchrotron radiation
from a mass $m$ moving in a circular orbit with period
$\omega_{0}$.  The power is \cite{Breuer} \eqn{syncis}{ P \sim {G
m^2 \omega_{0}^2 \gamma^4} \sim {G m^2 \over{b^2}} \gamma^4} where
we have identified the inverse frequency with the minimum expected orbital
radius, which is of order the impact parameter.  This result will
be very useful for understanding the decay of the initial circular
orbit.

Furthermore, one can directly compute, in the supergravity limit, the
radiation from an accelerated D-brane.  The result for circular motion
with radius $b$ is \cite{Costa} \eqn{radis}{ P = {G m^2\over{b^2}}
\gamma^4} The results \eno{syncis},\eno{bremisnow}, and \eno{radis} are
thus in good agreement.

Knowing now the power lost to closed strings for a given
decelerating trajectory, we also wish to compute the quantity of
open strings which would be required to produce this trajectory.
Stated more generally, given an object being accelerated by an
external force, we are interested in the ratio of the radiated
power to the power associated with the driving force.  For an
accelerating electron this is a textbook problem; see e.g.
\cite{Jackson}, chapter 14.

The result is that there is a characteristic length $L_{e} =
{2\over{3}}{e^2\over{mc^2}}$ governing radiation by electrons, and
unless an electron's energy changes by of order its rest energy
during acceleration over a distance of order $L_{e}$, the
radiation is negligible compared to the external power.  More
specifically, \eqn{stopelec}{ {E_{radiated}\over{E_{driving}}}
\equiv \Omega_{e} \approx{\Delta E\over{\Delta x}}{L_{e}\over{m
c^2}}} where the total change in energy, from all causes, is
$\Delta E$ over a distance $\Delta x$.

One can readily estimate the corresponding characteristic length
$L_{D}$ for massless closed string radiation from a D-brane by
comparing to the power \eno{radis}.  The outcome is that $ L_{D}
\sim g_{s} l_{s} $.

Let us now consider a brane whose initial kinetic energy is
$E_{i} = \gamma_{i} T_{p}$, where $\gamma_{i} \gg 1.$
Suppose that the brane decelerates over a
distance $\Delta x$ to a new kinetic energy $E_{f} = \gamma_{f} T_{p}$,
$\Delta \gamma \equiv \gamma_{i}-\gamma_{f}$. The `driving force'
here is loss of energy through open string production; we will now
compare this to the energy lost to radiation. \eqn{omegad}{
\Omega_{D} \equiv {E_{closed}\over{E_{open}}} \approx  {\Delta
E\over{T_{p}}}{L_{D}\over{\Delta x}} = g_{s} \Delta \gamma
{l_{s}\over{\Delta{x}}}}

If $\Omega_{D} \ll 1$ then our previous conclusions hold
automatically, as the closed strings are energetically negligible.
If $\Omega_{D} \gg 1$, there are two cases to consider.  First, if
$\gamma_{f} \sim 1$, so that $\Delta \gamma \sim \gamma_{i} \gg
1$, the branes have slowed down to nonrelativistic motion.  In
this case the energy in open strings can be estimated to be
\eqn{eopen}{ E_{open} \approx {\Delta E \over{\Omega_{D}}} \approx
{T_{p}\over{g_s}} {\Delta x\over{ l_s}}.}  To arrive at this rough
estimate we did not need the Bogoliubov coefficients derived from
the annulus amplitude; we have used instead the fact that the
external driving force (open string production) can be determined
based on the postulated trajectory.  Proceeding to estimate the
stopping length, we find \eqn{phiest}{{r_{*}\over{l_s}} \approx
{2 \pi \eta E_{f}\over{\nu_{open}}} \approx {\sqrt{2}\eta E_{f}\over{E_{open}}} \ll {\eta T_{p}\over{E_{open}}} = \eta g_{s}
{l_{s}\over{\Delta x}}.}  The distance $\Delta x$ is of order $\eta$, because that is the size of the nonadiabatic region
 in which open strings are created.  To make a very conservative estimate, we will use $\Delta x \gsim l_{s}.$  Then, because we are working at weak string coupling,
the stopping length is \eqn{risnow}{ {r_{*}\over{l_{s}}} \ll \eta g_{s} {l_{s} \over \Delta x} \ll \eta}
so that the stopping length is much smaller than $\eta l_{s}$.

The second case is $\Omega_{D} \gg 1, \gamma_{f} \gg 1$, so that the
brane is moving relativistically even after decelerating, and the
relative velocity is large when the branes pass each other. Our
general conclusion will be invalid only if the branes do not
rapidly trap in this final case.  However, if the branes separate
to a considerable distance while moving rapidly, our annulus
amplitude computation of open string production applies directly.
In other words, by assuming that the branes can separate, we are
arranging that they leave the region of nonadiabaticity, so that
the number density of open strings is accurately given by
\eno{FastProd}, and the trapping length by \eno{stopis}. Thus, the
assumption that the branes separate at high speed is not
consistent.

We conclude that closed string emission can slow the motion of the
brane pair, but it does not substantially increase the stopping
length. In fact, radiation helps considerably to bring the branes
to rest: once the branes are trapped and are spiraling around each
other, rapid radiation losses will slow their rotation.  This is
enhanced by the familiar fact that, for relativistic objects,
radiation losses are greater in circular motion than in
rectilinear accelerated motion.  Once the branes are trapped they
slow down through this closed string synchrotron radiation.  From
the power \eno{syncis} we conclude that the branes lose energy so
rapidly that they would require only a few orbits to come to rest.
In practice the spin-down process is prohibitively complicated, but this
result suffices to show that the lifetime of the highly-excited,
rapidly revolving remnant is in any case very short.

One important additional point is that the closed string radiation
is strictly negligible only when the coupling is so small that the
branes are rather heavy, and hence stretch the open strings farther before stopping.
There is consequently a tradeoff between computability and control, which are best at extremely weak coupling, and the
strength of the trapping, which is best for couplings above the bound \eno{gsis}.  It is essential to recognize that for
any nonzero coupling, the collision is inelastic and trapping eventually does occur; however, the stopping length increases
when the coupling grows very small.

A further question which we have not addressed is the production
of massive closed strings.  In the case of very abrupt
deceleration we would expect nonvanishing production of these
modes.  We will leave a precise computation of this effect within
string theory as an interesting problem for future work.

For the present analysis, we can make a very crude estimate of
massive string production by using a result on the spectrum of
gravitational synchrotron radiation.  For a mass in an orbit with
period $\omega_{0}$, the power per unit frequency is \cite{Breuer}
\eqn{freqspec}{ {{d P} \over{d \omega}} \propto \exp\Bigl(-{
\omega\over{\omega_{crit}}}\Bigr)} where $\omega_{crit} =
{6\over{\pi}}\gamma^2 \omega_{0}.$ Thus, for $\omega_{crit}
\ll {l_{s}^{-1}}$, massive closed strings should play a negligible role,
but
when gravitons of frequency $l_{s}^{-1}$ are being produced, it is
natural to expect massive modes as well.  We therefore expect some
emission of massive closed strings in processes where $ \gamma^2
\gg {b\over{l_{s}}}.$  This will further increase the rate of energy loss
from the revolving brane pair, speeding the trapping and increasing the effective inelasticity of the collision.

\subsection{ Summary of the Argument }

For clarity, we will now briefly review our argument that the
trapping of relativistic D-branes is powerful and abrupt.

The annulus partition function for open strings between moving
D-branes indicates that the density of produced open strings is
given, in the relativistic limit, by \eno{FastProd}.  The
characteristic impact parameter below which the backreaction of these
strings is strong can then be seen to be $b_{crit} \sim \eta
l_{s}$.  This is also the size $r_{nad}$ of the region in which the open
string masses change nonadiabatically.

If the D-branes are assumed to separate to a distance larger than
of order $b_{crit}$, they have left the region of nonadiabaticity,
so that \eno{FastProd} applies.  The energy \eno{eeta} in
open strings then exceeds the initial brane energy, so that the
assumption of significant separation was inconsistent.

The same argument applies when closed string radiation is taken
into account.  A straightforward estimate of the energy lost to
radiation over a distance $b_{crit}$ shows that the energy
transferred to open strings is still sufficient to stop the branes
before their separation exceeds $b_{crit}$.

We expect that a detailed computation of the string production along a
decelerating trajectory would show that the stopping length is at most of
order $b$, which can be much smaller than $b_{crit}$.  In particular, we
expect that in a head-on collision with negligible impact parameter the
stopping length would be of order the string length.  However, estimates
involving \eno{FastProd} are strictly valid only when the branes
eventually leave the window of nonadiabaticity, leading to the {\it{very}}
conservative estimate $r_{*} \sim b_{crit} = \eta l_{s} $.

A few potential objections remain.  First of all, one might worry
that the branes somehow slow down before reaching each other, so
that at the moment of closest approach the velocities are
nonrelativistic.  In this case excited open strings would not be
produced and we would simply have field theory trapping.  We have
already explained in \S4.3 that if the branes slow down
exclusively due to open string production, then they will still
experience rapid trapping.  Then, we showed in \S5.1 that
additional loss of energy through closed string radiation also
does not ruin the trapping.

A final worry is that the branes could interact by creating string
pairs at extremely high excitation levels. A vanishingly small number
density of arbitrarily highly excited strings (with level much higher
than $\eta^2$) could absorb all the initial kinetic energy and yet not
generate a strong attractive force between the branes. However, we
have seen that in fact string production peaks around level $n_{peak}
\approx {\eta^2}{(2\pi^2)}^{-1}$, which is sufficiently small to ensure
that the trapping is strong.

We therefore conclude that D-branes in relativistic motion
generically trap each other through copious production of open
strings, with a trapping length no larger than the size $\eta
l_{s}$ of the nonadiabatic region.  A sizeable fraction of the
initial energy is eventually emitted in the form of massless
closed string radiation.

The limitations to our argument which we have discussed above
make it challenging to precisely and controllably compute the stopping length in an ultrarelativistic D-brane
collision.  However, these issues, and others -- such as massless and massive closed string radiation, annihilation of the produced strings,
and dilution of the produced strings in a cosmological background -- do not in any way weaken our argument that the brane collision is inelastic.
In fact, it is easy to see that radiation, annihilation, and dilution all extract energy from the brane system, slowing the brane
motion.  (See \cite{KLLMMS} for an analysis of these issues in the nonrelativistic context.)  Happily, for applications to cosmological models, it is the inelasticity rather than the stopping length which is most immediately relevant.

\subsection{ Regime of Validity and Control }

We will now examine the characteristics of the trapping process as
 a function of the dimensionless parameters $g_{s}, b, \eta$.

First of all, we will never work at strong string coupling ($g_s > 1$),
since then we would have to include higher string loop effects.
Furthermore, at strong coupling the D-branes become very light,
and their Compton wavelength $\lambda_{D}$ grows.  We require
$\lambda_{D} \ll b$ so that we can neglect these quantum effects.

Secondly, we should require that the Schwarzschild radius $R_{s}$ of
the D-brane is negligibly small compared to $b$.  To estimate this, we
treat the $Dp$-brane as a point source in
$10-p$ dimensions. The black hole solution in $(10-p)$ dimensions
for a p-dimensional extended object of tension T and zero charge
is \cite{BBrane}\ \eqn{rs}{ T =
{\left({{8-p}\over{7-p}}\right)}{R_{s}^{7-p}\over{(2\pi)^7 d_{p}
g_s^2}} } where $d_{p} =
2^{5-p}\pi^{{5-p}\over{2}}\Gamma\left({{7-p}\over{2}}\right).$

We are interested in the limit of zero charge because the
highly-boosted branes have far greater effective mass than the BPS
bound requires.  Note that in fact the metric for one of these
moving branes is of a shock-wave form, not a static black hole.  We
are imagining that the branes collide inelastically and then
asking whether the Schwarzschild radius of the excited remnant,
seen in the center of mass frame, is comparable to the initial
impact parameter.

In this scheme, the effective tension is the center-of-mass energy
$2 T_{p} \gamma \approx  T_{p} e^{\eta/2}$. We therefore find,
using the tension of a $p$-brane, \eqn{rsis}{ \left({R_{s}\over{
l_{s}}}\right)^{7-p} = g_s {\left({{7-p}\over{8-p}}\right)}
(2\pi)^{7-p}  d_{p} e^{\eta/2} } from which we conclude that for
$p<7$, the
Schwarzschild radius can be made parametrically less than any
given impact parameter by reducing the string coupling.

Let us now fix $b$ and $\eta$ and take the string coupling to be
small enough so that string loops, the brane Schwarzschild radius, and the brane
Compton wavelength can be neglected.  As we further decrease the
coupling, the brane becomes heavier and the stopping length
becomes greater.  Now, recall that when we examined the open
string production along a constant-velocity path, we found an
energetic inconsistency: unless the coupling was exponentially
small, the open string energy exceeded the initial kinetic energy
of the system.  Of course, deceleration reduces string production,
so for any controllable coupling the energy in open
strings will not exceed the initial energy.  However, we can still
define a value of $g_{s}$ at which the energetics is consistent
even before we incorporate the deceleration which arises from
backreaction. Comparing \eno{energetics} and \eno{open}, we find
that the energetics are automatically consistent provided that
\eqn{consist}{ g_{s} < 2^{3/2} {\pi}^{p/2-3} \eta^{4-p/2} e^{-\eta/2}. }
Thus, only for exponentially small string coupling are the branes
so heavy that they stretch the open strings substantially before
coming to rest.

\section{Discussion}

We have argued that the relativistic scattering of Dp-branes,
$p>0$, at small impact parameters is almost completely inelastic
as a result of pair production of excited open strings. The
time-dependence induces production of an extremely high density of
highly-excited, stretched open strings, which rapidly draw the
branes into a tight orbit whose radius is of order the impact
parameter.  The resulting acceleration results in significant
closed string radiation, which acts to further brake the motion.

Powerful stringy trapping of this sort occurs whenever the impact
parameter, measured in string units, is small compared to the
rapidity $\eta.$  This is a much larger range of distances than
that controlled by collision of the stringy halos of the two
branes, whose radius grows as $\sqrt{\eta}$.  Moreover,
the strength of this stringy trapping was a surprise: it does not
follow from summing the low-velocity result of \cite{KLLMMS} over
the string spectrum.  Instead, the velocity-dependence of
stretched string masses enters in a crucial way to enhance the
production effect.

Our result, which is essentially a simple observation about the
quantum-corrected dynamics of D-branes, has obvious implications
for scenarios involving branes in relativistic motion.  One example\footnote{We are grateful to S. Kachru
for suggesting this.} is the stage of reheating in
cosmological models with fast-moving branes and antibranes.
Brane-antibrane inflation models typically end with the
condensation of the open string tachyon, leaving a dust of closed
strings in the bulk as well as excited open strings on any
remaining branes \cite{Sen}. Despite much effort, this process is
not fully understood \cite{LambertLM}.  Suppose, however, that the
antibrane is moving relativistically toward the end of its
evolution, and then passes by or collides with a stack of branes.
(Ultrarelativistic brane motion is natural in the DBI models
\cite{Dccel,DBI}, for example, and could occur elsewhere.) In this
case tachyon condensation governs only a small fraction of the
energy released; most of the kinetic energy goes into open string
pair production. Thus, reheating in such a model proceeds by
stringy trapping (for related work, see \cite{Chen}).

More speculatively, moduli trapping may be a useful mechanism for
vacuum selection \cite{KLLMMS}, as it gives a dynamical explanation
for the presence of enhanced symmetry.  (See also \cite{Watson,JarvMS} for related work on moduli dynamics
in string/M theory.)   The stringy trapping
presented here extends the trapping proposal not just to a new
parameter range, but to a regime where the strength of the effect
increases dramatically.

The inelasticity of D-brane scattering may be viewed as a
calculable example of a more general question: to what extent do
particle, string, and brane production affect motion toward or
away from a given `singular' configuration? Time-dependent
orbifolds
\cite{Liuone,Liutwo,Horowitz,Fabingermc,Fabingerhell,Pioline,Berkooz}
(see also \cite{Cornalba:2003kd} and references therein) provide a
relatively tractable setting for such a question. Berkooz and
Pioline \cite{Pioline} and Berkooz, Pioline and Rozali
\cite{Berkooz} have emphasized the possibility of resolving a
spacelike singularity through the pair production and condensation
of winding strings. It would be very interesting to extend these
results and repair more general spacelike singularities through
the production of branes or strings; see \cite{Friess} for work in
this direction. Our analysis suggests that string production could
be surprisingly important in such a setting.

Another interesting open question is whether the inelasticity of
quantum-corrected D-brane collisions can be used to place bounds
on the elasticity of other sorts of collisions. In the cyclic
universe model \cite{Khoury,Cyclic}, the orbifold boundaries of heterotic
M-theory \cite{HW} approach each other and collide. An intrinsic
assumption of these cyclic models is that the collision is very
nearly elastic; this is essential to make possible a large number
of collisions and the associated cyclic behavior. Our result makes
it plain that D-brane collisions, which appear elastic
classically, are highly inelastic when the quantum effects
associated to fundamental strings are included.

In the cyclic model, the M2-branes stretched between the
boundaries become tensionless at the instant of collision.  In the
weakly-coupled four-dimensional description these objects are
heterotic strings whose tension, in four-dimensional Planck units,
goes to zero at the moment of impact. Because the masses vary
rapidly during the collision, the nonadiabaticity is large and we
expect copious production of these strings.  It would be extremely
interesting to compute the energy loss through this
string/membrane production and to understand the implications for
the cyclic models \cite{pjs}.

We should point out that in the most realistic cyclic models, the
brane velocities are required, for phenomenological reasons, to be
nonrelativistic.\footnote{We are grateful to P.~J.~Steinhardt for
helpful discussions on this point.}  The results in this paper
appear to give an independent upper bound on the velocity of the
branes before collision -- this bound is one which is required for
the self-consistency of the model, rather than one imposed by
observational requirements. However, this argument is qualitative
at present; an explicit extension of our results involving
stretched fundamental strings to the case of stretched membranes
would be nontrivial.
 
 Another interesting application would be to investigate inelasticity
 in the relativistic dynamics of networks of cosmic strings \cite{JJP}. 

\section*{Acknowledgements}

We are very grateful to E.~Silverstein for collaboration in the
early stages of this project and for helpful comments throughout.
We would also like to thank R.~Bousso, M.~Fabinger, B.~Freivogel, O.~Ganor, S.~Giddings, S.~Gubser,
P.~Horava, S.~Kachru, X.~Liu, A.~Maloney, M.~M.
~Sheikh-Jabbari, S.~Shenker, and P.~J.~Steinhardt for helpful discussions, and S.~Gubser, E.~Silverstein, and P.~J.~Steinhardt
for comments on the manuscript.  L.M. was supported in part by the Department of Energy under contract number
DE-AC03-76SF00515, and is grateful to the organizers of the 2004 PiTP program in String Theory, where a portion of this work was done.  I.M. was
supported in part by the Director, Office of Science, Office of High
Energy and Nuclear Physics, of the U.S. Department of Energy under
Contract ~DE-AC03-76SF00098, and in part by the NSF under grant
PHY-0098840 and grant PHY-0244900.  I.M. also wishes to thank the Aspen
Center for Physics for hospitality during the final stages of this work.

\newpage

\section*{Appendix}

 In this section we collect various identities about the elliptic
 theta functions. Because of the existence of several canonical notations for
 these functions, we define the functions as used in the paper.

 The theta functions are often expressed in terms of the variables
 $\nu$ and $\tau$, or in terms of the nome $q= \exp(2 \pi i \tau)$ and
 $z= \exp (2 \pi i \nu)$. The four theta functions are written down
 below in both their series and product forms:

\eqn{Theta}{\eqalign{
 \theta_{00} (\nu, \tau)  = \theta_3 (\nu | \tau)  & =
 \sum_{n=-\infty}^{n=\infty} q^{n^2/2} z^n
 = \prod_{m=1}^{\infty} (1-q^m)(1+zq^{m-1/2})(1+z^{-1}q^{m-1/2}) \cr
 \theta_{01} (\nu, \tau)  = \theta_4 (\nu | \tau)  & =
 \sum_{n=-\infty}^{n=\infty} (-1)^n q^{n^2/2} z^n
= \prod_{m=1}^{\infty} (1-q^m)(1-zq^{m-1/2})(1-z^{-1}q^{m-1/2})
\cr
   \theta_{10} (\nu, \tau) = \theta_2 (\nu | \tau) & =
 \sum_{n=-\infty}^{n=\infty} q^{(n-1/2)^2/2} z^{n-1/2} \cr
& = 2 e^{\pi i \tau/4} \cos(\pi \nu) \prod_{m=1}^{\infty}
 (1-q^m)(1+zq^m)(1+z^{-1}q^m) \cr
 - \theta_{11} (\nu, \tau) = \theta_1 (\nu | \tau) & = i
 \sum_{n=-\infty}^{n=\infty} (-1)^n q^{(n-1/2)^2/2} z^{n-1/2} \cr
 & = 2 e^{\pi i \tau/4} \sin(\pi \nu) \prod_{m=1}^{\infty}
 (1-q^m)(1-zq^m)(1-z^{-1}q^m) \,.
 }}

 In addition to the theta functions, we shall also need the Dedekind eta
 function:
\eqn{Eta}{
 \eta (\tau) = q^{1/{24}} \prod_{m=1}^{\infty} (1 - q^m) = \left[
 {{\partial_\nu \theta_{11} (0, \tau)} \over {-2 \pi}} \right]^{1/3} \,.
}

These functions have the following modular transformation properties:
\eqn{Modular}{\eqalign{
 & \theta_{00}(\nu /\tau, -1/\tau)= (-i \tau)^{1/2} \exp(\pi i
 \nu^2/\tau) \theta_{00}(\nu, \tau) \cr
 & \theta_{01}(\nu /\tau, -1/\tau)= (-i \tau)^{1/2} \exp(\pi i
 \nu^2/\tau) \theta_{10}(\nu, \tau) \cr
 & \theta_{10}(\nu /\tau, -1/\tau)= (-i \tau)^{1/2} \exp(\pi i
 \nu^2/\tau) \theta_{01}(\nu, \tau) \cr
 & \theta_{11}(\nu /\tau, -1/\tau)= - (-i \tau)^{1/2} \exp(\pi i
 \nu^2/\tau) \theta_{11}(\nu, \tau) \cr
 & \eta(-1/\tau) = (-i \tau)^{1/2} \eta(\tau) \,.
}}

We will often need the asymptotic behavior of the theta and eta
functions. When $q \ll 1$ we can immediately find the asymptotics using the above
expansions, whereas for $q\to 1$ we must first perform a modular
transformation.

The asymptotic behavior of a particular combination will be especially helpful.
Define the fermionic partition function $Z(\tau) \equiv  {1\over{2}}\theta_{10}^4(0|\tau)\eta(\tau)^{-12}.$
Then for $-i\tau \equiv s \gg 1$ we have
\eqn{otherasympt}{ Z(i s) = 8 + {\cal{O}}(e^{-2 \pi s})}
%{ Z(i s) = 2 \exp\Bigl({3\pi\over{4}}s\Bigr)\Bigl(1 + {\cal{O}}(e^{-2 \pi s})\Bigr)}
whereas for $ s \ll 1$ we find, using the modular transformations above,
\eqn{asympt}{ Z(i s) = {1\over{2}}s^4 \exp\Bigl({\pi\over{s}}\Bigr)\Bigl(1 + {\cal{O}}(e^{-{\pi\over{s}}})\Bigr)}

We will also need a few identities involving the theta functions:
\eqn{ThetaIden}{\eqalign{
& \theta_{00}^4 (0,\tau) - \theta^4_{01}(0,\tau) - \theta^4_{10}
 (0,\tau) = 0 \qquad
 \theta_{11} (0,\tau) = 0 \cr
& \prod_{a=1}^4 Z^0_0 (\phi_a, it) - \prod_{a=1}^4 Z^0_1 (\phi_a, it)
- \prod_{a=1}^4 Z^1_0 (\phi_a, it) - \prod_{a=1}^4 Z^1_1 (\phi_a, it)
= 2 \prod_{a=1}^4 Z^1_1 (\phi'_a, it) \,,
}}
where
\eqn{phidef}{\eqalign{
& Z^{\alpha}_{\beta} (\phi, it) = {\theta_{\alpha \beta}(i \phi t/\pi,
  it) \over {\exp(\phi^2 t/\pi) \eta(it)}} \cr
& \phi'_1 = {1 \over 2}(\phi_1 + \phi_2 + \phi_3 + \phi_4) \qquad
  \phi'_2 = {1 \over 2}(\phi_1 + \phi_2 - \phi_3 - \phi_4) \cr
& \phi'_3 = {1 \over 2}(\phi_1 - \phi_2 + \phi_3 - \phi_4) \qquad
  \phi'_4 = {1 \over 2}(\phi_1 - \phi_2 - \phi_3 + \phi_4) \,.
}}
The identity \ThetaIden\ leads in the case $\phi_2=\phi_3=\phi_4=0$ to
\eqn{secondident}{2 \theta_{11}^4(\nu/2,\tau)
=\theta_{00}(\nu,\tau)\theta_{00}^3(0,\tau)-
\theta_{01}(\nu,\tau)\theta_{01}^3(0,\tau)-
\theta_{10}(\nu,\tau)\theta_{10}^3(0,\tau).}

\bibliographystyle{ssg}
\bibliography{0810}

\end{document}